\input harvmac

\def\Title#1#2{\rightline{#1}\ifx\answ\bigans\nopagenumbers\pageno0
\vskip0.5in
\else\pageno1\vskip.5in\fi \centerline{\titlefont #2}\vskip .3in}

\font\caps=cmcsc10

\noblackbox
\parskip=1.5mm

  
\def\npb#1#2#3{{\it Nucl. Phys.} {\bf B#1} (#2) #3 }
\def\plb#1#2#3{{\it Phys. Lett.} {\bf B#1} (#2) #3 }
\def\prd#1#2#3{{\it Phys. Rev. } {\bf D#1} (#2) #3 }
\def\prl#1#2#3{{\it Phys. Rev. Lett.} {\bf #1} (#2) #3 }
\def\mpla#1#2#3{{\it Mod. Phys. Lett.} {\bf A#1} (#2) #3 }
\def\ijmpa#1#2#3{{\it Int. J. Mod. Phys.} {\bf A#1} (#2) #3 }

\def\cmp#1#2#3{{\it Commun. Math. Phys.} {\bf #1} (#2) #3 }

\def\ptp#1#2#3{{\it Prog. Theor. Phys.} {\bf #1} (#2) #3 }


\def\half{{1\over 2}\,}

\def\pt{\partial}
 

            \def\CZ{{\cal Z}}
  \def\CF{{\cal F}} 
 \def\CH{{\cal H}}

 \def\CS{{\cal S}}  

\lref\rade{M. Ademolo, L. Brink, A. D'Adda, R. D'Auria, E. Napolitano,
S. Sciuto, E. Del Giudice, P. Di Vecchia, S. Ferrara, F. Gliozzi, R.
Musto and R. Pettorino, J.H. Schwarz, \npb {111}{1976}{77.}} 
\lref\rgreen{M. B. Green, \npb {293}{1987}{593.}}
\lref\roov{H. Ooguri and C. Vafa, \mpla {5}{1990}{1389;} \npb
{361}{1991}{469;} \npb {367}{1991}{83.}} 
\lref\rwit{E. Witten, \prl {61}{1988}{670.}}
\lref\rgrossm{D.J. Gross and P.F. Mende, \plb {197}{1987}{129;}
 \npb {303}{1988}{407.}}
\lref\rgrossmanes{D.J. Gross and J.L. Ma{\~n}es, \npb {326}{1989}{73.}}
\lref\rtria{R. Dijkgraaf, E. Verlinde and H. Verlinde in: {\it 
Perspectives in String Theory}, Proc. Copenhagen 1987,  World Scientific,
Singapore 1988 \semi
A. Giveon, N. Malkin and E. Rabinovici. \plb {220}{1989}{551.}} 
\lref\rkutsei{D. Kutasov and N. Seiberg, \npb {358}{1991}{600.}}
\lref\rrabino{A. Giveon, N. Malkin and E. Rabinovici, \plb
{220}{1989}{551.}}
\lref\ratickw{J. Atick and E. Witten, \npb {310}{1988}{291.}}
\lref\rbranva{R.H. Brandenberger and C. Vafa, \npb {316}{1988}{391.}}
\lref\rbekleb{ D.J. Gross and I.R. Klebanov, \npb
{344}{1990}{345\semi} 
M. Bershadsky and I.R. Klebanov, \prl {65}{1990}{3088.}}
\lref\rkutmar{D. Kutasov and E. Martinec, Preprint 
EFI-96-04, hep-th/9510182.} 
\lref\rkutmaro{D. Kutasov,
 E. Martinec and M. O'Loughlin, Preprint EFI-96-07, hep-th/9603116.}
\lref\rpol{J. Polchinski, \cmp {104}{1986}{37\semi} 
K.H. O'Brien and C.I. Tan, \prd {36}{1987}{1184\semi} 
B. McClain and B.D.B Roth, \cmp{111}{1987}{539.}}
\lref\rao{E. Alvarez and M.A.R. Osorio, \npb{304}{1988}{327.}}
\lref\rtdual{K. Kikkawa and M. Yamasaki, \plb{149}{1984}{257\semi}
N. Sakai and I. Senda, \ptp{75}{1984}{692\semi}
P. Ginsparg and C. Vafa, \npb{289}{1987}{914\semi}
E. Alvarez and M.A.R. Osorio, \prd{40}{1989}{1150.} }
\lref\rramp{R. Rohm, \npb{237}{1984}{553.}}
\lref\rorb{M.A.R. Osorio and M.A. V\'azquez-Mozo, \prd{47}{1993}{3411.}}
\lref\ros{M.A.R. Osorio, \ijmpa{7}{1992}{4275.}}
\lref\rmal{G. Moore, \npb{293}{1987}{139.}}
\lref\rdkl{L. Dixon, V. Kaplunovsky and J. Louis, \npb{355}{1991}
{649\semi} J. Harvey and G. Moore, Preprint EFI-95-64,
hep-th/9510182.}
\lref\rosva{M.A.R. Osorio and M.A. V\'azquez-Mozo, \plb{280}{1992}{21.}}
\lref\ras{M. Abramowitz and I.A. Stegun, {\it ``Handbook of Mathematical 
Functions''}, Dover, New York 1965.}
\lref\rsdual{E. Witten, \npb{443}{1995}{85\semi}
J. Polchinski, Preprint NSF-ITP-95-157, hep-th/9511157.}
\lref\rkoblitz{N. Koblitz, {\it Introduction to Elliptic Curves and Modular
Forms}, Springer, Berlin 1984.}
\lref\rc{M.A.R. Osorio and M.A. V\'azquez-Mozo, Preprint 
IASSNS-HEP-95-96 \semi
hep-th/9511157 ({\it Nucl. Phys. B}, to appear)\semi
D. Pierce, Preprint IFP-604-UNC, hep-th/9601125.}
\lref\rmm{S.D. Mathur and S. Mukhi, \npb{302}{1988}{130.}}
\lref\rnsv{K.S. Narain, M.H. Sarmadi and C. Vafa, \npb{288}{1987}{551.}}
\lref\rseipol{J. Dai, R.G. Leigh and J. Polchinski, \mpla
{4}{1989}{2073\semi}
M. Dine, P. Huet and N. Seiberg, \npb {322}{1989}{301.}}
\lref\rvene{D. Amati, M. Ciafaloni and G. Veneziano, \plb
{197}{1987}{81\semi}
\ijmpa {3}{1988}{1615.}}
\lref\rmende{P.F. Mende, \plb {326}{1994}{212.}}


\line{\hfill PUPT-96-1620}
\line{\hfill IASSNS-HEP-96-38}
\line{\hfill {\tt hep-th/9605050}}
\vskip 0.5cm

\Title{\vbox{\baselineskip 12pt\hbox{}
 }}
{\vbox {\centerline{Complex World-Sheets from $N=2$ Strings }
}}

\centerline{$\quad$ {\caps J. L. F. Barb\'on}}
\smallskip
\centerline{{\sl Joseph Henry Laboratories}}
\centerline{{\sl Princeton University}}
\centerline{{\sl Princeton, NJ 08544, U.S.A.}}
\centerline{{\tt barbon@puhep1.princeton.edu}}
\vskip 0.1in

\centerline{$\quad$ {\caps M. A. V\'azquez-Mozo}}
\smallskip
\centerline{{\sl School of Natural Sciences}}
\centerline{{\sl Institute of Advanced Study}}   
\centerline{{\sl Princeton, NJ 08540, U.S.A.}}
\centerline{{\tt vazquez@sns.ias.edu}}
 \vskip 0.3in

We study some properties of target space strings constructed from
(2,1) heterotic strings.
We argue that world-sheet complexification is a general property of the
bosonic  sector of such target  world-sheets.
 We give a target space
interpretation of this fact and relate it  
 to the non-gaussian nature of free String Field Theory.  
 We provide several one loop calculations supporting the stringy
construction of critical world-sheets in terms of (2,1) models. 
Using finite temperature boundary
conditions in the underlying (2,1) string   we obtain non-chiral
target space spin structures, and point out some of the problems  
arising for chiral spin structures, such as the heterotic world-sheet.
To this end, we study the torus partition function of the
corresponding asymmetric orbifold of the (2,1) string.


\Date{5/96}


\newsec{Introduction}

String Theory, in its present perturbative 
formulation, is defined in terms of its
underlying world-sheet two dimensional field theory. However, recent
developments in non-perturbative string dynamics (see for example 
\refs\rsdual)  
seem to suggest that
the world-sheet description plays no such
 fundamental r\^ole,  
 and it is only valid 
in some corners of the moduli space of vacua. Then, any non-perturbative
formulation of String Theory must reduce to perturbative   world-sheet
 descriptions  
 in some weak or strong coupling
regimes. An interesting scenario, in which different world-sheet theories
 appear
to be derived from a more fundamental principle, was introduced by 
Green \refs\rgreen, who proposed to obtain string world-sheets
as target space theories of two-dimensional strings. In this way, space-time
coordinates arise dynamically as the massless excitations of some
underlying string theory living in a two dimensional target space. 
A serious difficulty  one meets in trying to recover 
strings from strings is the fact that  the massless modes
of the underlying string theory have complicated interactions induced
by the massive modes which have been integrated out. Another
problem is that, in general, the field content corresponds to a
non-critical rather than critical world-sheet.  

One way of keeping interactions between massless modes under control
is by considering $N=2$ strings \refs\rade\ as the underlying theory.
$N=2$ strings have a number of interesting features \refs\roov. Their critical
dimension is four but, because of the existence of $N=2$ superconformal 
symmetry, the target manifold must have a complex structure so
the target signature is either $4+0$ of $2+2$. The $N=2$ superconformal algebra
contains, in addition to the two fermionic generators $\overline{G}^{\pm}$, 
a $U(1)$ current $\overline{J}$. In heterotic constructions of the type 
$(2,0)$ 
or $(2,1)$ this current has to be balanced by a left-moving counterpart $J$ 
whose gauging forces the introduction of a new set of 
ghosts  raising the critical dimension in the left-moving sector 
by 2. In the case of the $(2,1)$ models this means that we have a left-moving 
internal $N=1$ SCFT with $\hat{c}=8$. Absence of BRST anomalies further 
imposes the left-moving $U(1)$ current $J$ has to lie in a
null subspace. Finally, their spectrum is very simple, containing only a
finite number of massless modes,  and their $n$-point scattering amplitudes 
vanish for $n\geq 4$. 
For the $(2,1)$ models with two-dimensional target space, even the on-shell
three-point functions vanish.

Recently, Kutasov and Martinec \refs\rkutmar\ have rescued the idea of Green 
and proposed that both string world-sheets and membrane world-volumes
can be obtained as different vacua of $(2,1)$ heterotic strings. Taking
$J$ to have no component on the left-moving internal sector
the string theory lives effectively in $1+1$ dimensions and 
one can recover the classic   bosonic, type-IIB
and heterotic world-sheet theories in a physical gauge, 
 for different choices of the internal $\hat{c}=8$ 
SCFT. By relaxing the condition that $J$ 
has to lie entirely in the $2+2$ non compact space one gets an effective 
$2+1$ theory that corresponds to the world-volume theory of two dimensional 
membranes. 
In fact, the $(2,1)$ string construction seems to provide a unified
picture of all ``M-brane" vacua \refs\rkutmaro.

The computation of the torus partition function in the target world-sheet 
theory involves now the evaluation of string vacuum amplitudes in the
underlying theory. In fact, the results obtained at one loop order in the
topological expansion of the $N=2$ string suggest that there are no higher 
order corrections in $\lambda$, the $N=2$ string coupling constant. There 
are however non-perturbative corrections
of order $e^{-1/\lambda^{2}}$ which would account, for example, for
global features of the target world-sheet fields.
In this paper we will only consider one loop partition functions both in 
the $N=2$ and in the target string sense, thereby summing embeddings of 
the world-sheet tori onto target two dimensional tori.

Since we are dealing with a string theory in a toroidal space-time 
we will have winding as well as momentum modes around,  and  the first 
impression is that we will get twice the number of states we would like to.
This fact has been noticed in the  case of $N=2$ strings  
\refs\roov\rdkl\ from the following peculiar form of
the one-loop free energy in a target torus with modular parameters $T$ and 
$U$:
\eqn\uno
{{\rm log}\,
 Z(T,U)_{\rm target} = -D_T \, {\rm log}\, \left( \sqrt{T_2} |\eta (T)|^2
     \,\,\sqrt{U_2} |\eta (U)|^2 \right), }
where $D_T$ is the effective 
number of transverse degrees of freedom (number of target space free
bosons minus fermions). For the $(2,2)$ 
string $D_T = 2$, whereas for the $(2,1)$ string whose target space dynamics
is the standard bosonic world-sheet, $D_T = 24$. In general, $D_T =0$
for supersymmetric strings with vanishing two-dimensional cosmological
constant. $T$ and $U$ are respectively the complex and the K\"ahler moduli
of the target torus, defined in terms of the background values of the 
metric and the torsion as
$$
T = {g_{12}+i\sqrt{\det g}\over g_{22}} \hskip 1cm
\alpha' U = b_{12}+i\sqrt{\det g}.  
$$
One way of getting rid of the unwanted modes is just by taking the 
field theory limit in the underlying theory, decoupling all winding
modes by sending  the string tension to 
infinity or, equivalently, by stretching the target 
world-sheet to infinite
area. However, as it
stands, the most striking feature of \uno\ is that the number of
massless degrees of freedom in the target space appears doubled, as well
as the target space moduli parameters. The complex structure moduli $T$
enters as a standard world-sheet moduli parameter and, remarkably,  
 the K{\"a}hler parameter $U$ is reinterpreted as the 
complex structure of an identical decoupled world-sheet theory\foot{The
$T$-$U$ symmetry in \uno\ is related to the stronger ``triality"
symmetry of the integrand of $N=2$ partition functions. This symmetry
is characteristic of the solitonic sector \refs\rtria.}. Thus, the
effective spacetime of the target theory is a complex world-sheet, with
the ``imaginary" component being provided by the mirror pair.

Complex
world-sheets have been claimed to be involved in the description of
the high-energy phase of String Theory, either from the study of 
strings at finite temperature \refs\ratickw\ or from the behavior
of string scattering amplitudes at high energies \refs\rgrossm\rwit.
It is then natural to consider \uno\ as an indication that the target
string construction based on underlying $N=2$ models is in some sense
related to the high energy phase of standard strings. The relation of
the $N=2$ construction to M-theory \refs\rkutmaro\ makes this
observation very suggestive.

The relevance of mirror symmetry to the $T$-$U$ factorization phenomenon
suggests an $N=2$ world-sheet mechanism based on the properties of the
corresponding chiral rings (see \refs\rkutmar). In this paper we will
adopt a target space point of view and try to find the conditions for
factorization in a general  two-dimensional string background. 
The main conclusion is that the 
target space of two dimensional string theories, as defined by
massless {\it bosonic}  
probes, is generally complexified. Interpreted as a world-sheet 
theory itself, we are led to a theory with complex world-sheets. Although
it is clear that the doubling of degrees of freedom has to do with
the winding modes of the underlying $(2,1)$ model, the precise
space-time interpretation of  
the factorization between $T$ and $U$ dependence is
not so obvious. By providing a physical interpretation of the
calculation of \uno\ in \rdkl\ (along the lines of de second reference
in \roov),
 we argue that such factorization is a rather general
feature of the massless sector of two dimensional String Field Theory,
and involves some simple, yet subtle properties of one-loop string path
integrals. We also find that the universality of the factorization is
lost when considering the fermionic sector of the target strings, even
in the simplest $(2,1)$ models.

The plan of the paper is as follows. In section 2 we will study in detail
the stringy mechanism behind the doubling of the target partition function
of the (2,1) models when all the target space fields are periodic. To this 
end we will analyze the correspondence between the massless sector of
two dimensional  String Field Theory 
and one loop computations in the first-quantized underlying  string
model. In section 3
we will be concerned with the study of target space spin structures from
the point of view of the underlying (2,1) string. We will compute the
target space partition function in the four sectors of non-chiral boundary
conditions of the target fermions of the type-IIB superstring and will 
outline some ideas of how to deal with chiral target spin structures. As 
an application of this proposal we will briefly study the case of the 
target $SO(32)$ heterotic string. Finally in section 4 we will summarize
 our
conclusions. The details of the computations in section 3 are given
in Appendix A, while  Appendix B contains a review  of 
$\beta$-duality.

\newsec{Fields vs. Strings on the target world-sheet} 

One of the most interesting 
distinctive features of $(2,1)$ strings is the simplicity
of the corresponding String Field Theory (SFT). Indeed, since the
spectrum consists of a finite number of massless free fields,
one 
is tempted to assume that $(2,1)$ SFT is more or less equivalent to free
Field Theory in the two-dimensional target space, once the conformal
gravitational factor and the dilaton have been gauged away. 
It is on this basis that one identifies the different critical
world-sheet theories as vacua of the $(2,1)$ models. However, the
doubling of degrees of freedom exhibited in \uno\ is an indication of
certain ``stringy" features which make the target world-sheet rather
exotic. In this section we present a general discussion of the formula
\uno, emphasizing the differences between SFT and standard Field
Theory. 

It will be instructive to keep the discussion as general as possible,
considering arbitrary string vacua with two dimensional space-time,
and concentrate on the free approximation. The conformal structure on
left and right movers depends on the gauged world-sheet gravity. In the
$N=0$ case we have a ``longitudinal" $c_L=2$ free CFT for the
space-time directions, with Minkowskian or Euclidean signature, and an
internal or ``transverse" $c_T = 24$ CFT. With $N=1$ worldsheet
supergravity we have a ${\hat c}_L = 2$ longitudinal SCFT, and   
a ${\hat c}_T = 8$ internal SCFT
($N=1$ SCFTs with $\hat{c}=8$ have been studied in \refs\rc). Finally, 
the $N=2$ case has a ${\hat
c} =4$ free system with $2+2$ signature,
 which is directly critical with only one complex
scalar degree of freedom. For the application to target string
theories, we can define models with two dimensional target space by
gauging a convenient null current. This procedure leaves no
propagating degrees of freedom, apart from the zero modes. So, the
$(2,*)$ strings have a purely holomorphic torus 
partition function, except
from the contribution of the zero modes.  

In a  $1+1$ Minkowski target space, the on-shell condition  is given by
\eqn\tres
{(L_0 + {\overline L}_0)\Psi_f  ={\alpha' \over 2}
 \left( p^2 + M^2 \right)
\Psi_f =0,}
where 
\eqn\cuatro
{M^2 ={2\over \alpha'} (L'_0 + {\overline L}'_0 )_L 
+ {2\over \alpha'} (L_0 + {\overline L}_0 )_T 
+ {2\over \alpha'} (L_0 + {\overline L}_0 )_{\rm ghost}  }
is the mass 
 operator, whose spectrum   is constrained by the various
physical state conditions, including appropriate 
GSO projections. In \cuatro\ we denote by $L'_0$ the oscillator
part of the ``longitudinal"  
 Virasoro operator.  
  Apart from global
(discrete) states, its excitations lead to longitudinal (gauge)
degrees of freedom.  

  At  the massless 
level one finds $D_T$ effective (physical) 
 degrees of freedom, counting bosons with
a $+1$ and fermions with a $-1$. Naively, the corresponding SFT action
can be written, in a somewhat symbolic form as, 
\eqn\dos
{S_{\rm target} = {1\over \alpha' \lambda^2}  \int_{\rm spacetime}
\sum_f \Psi_f \,(L_0 + {\overline L}_0)\, \Psi_f + {\rm
interactions,}}
where the two dimensional fields $\Psi_f$ have dimensions of length,
and $\lambda$ is the string coupling of the underlying string theory.
If \dos\ is interpreted as a target world-sheet action, the Regge
slope of the target string is given by $\alpha'_{\rm target} \sim
\alpha' \lambda^2 $. Accordingly, target world-sheet instanton effects
$\sim e^{-1/\alpha'_{\rm target}}$ correspond to non-perturbative
string effects in the underlying string model.

In an attempt to provide a target space explanation of \uno, we could
use the  action \dos\
 to compute the one-loop partition function in
the massless sector, when the space-time is an euclidean torus or
radii 
 $R_1$ and $R_2$. The result is 
\eqn\cinco
{{\rm log} \, Z_{\rm massless} \sim  
-{1\over 2} \, {\rm Tr}^{'}_{M^2 =0} \, (-1)^F \, {\rm log} \, (-\pt^2)
   \sim -D_T \, {\rm log} \, \left(
\sqrt{R_1 \over R_2} \,| \eta (iR_1 /R_2) |^2 \right) +{\rm constant.}  }   
where the prime stands for zero mode subtraction. 
For a straight torus with no torsion background we have $T=iR_1 /R_2$
and $U=iR_1 R_2$. 
Clearly, we fall short in reproducing \uno\ since we should get a second   
factor with $R_2$ replaced by $1/R_2$. 
This suggests that we need the winding
modes which are absent in \dos.      The diagonal kinetic kernel for
a target torus is in fact (in units $\alpha' =1$),  
\eqn\seis
{2(L_0 + {\overline L}_0 )  \equiv K_0 + M^2 = {n_1^2 \over 
R_1^2} + {n_2^2 \over R_2^2} + \ell_1^2 R_1^2 + \ell_2^2 R_2^2 + M^2. } 
The integers $n_1,n_2$ label momentum modes and $\ell_1,\ell_2$
 correspond to the
winding modes.
In a standard interpretation, we may regard the integers $\ell_1,\ell_2$
 as
labels of additional massive fields, with an effective mass 
$M^2_{\rm eff} = M^{2}+\ell_1^2 R_1^2 + \ell_2^2 R_2^2$. However, it is 
well known
that, when writing  \dos, there is  
 a very general ambiguity  in the ``definition" of 
space-time, versus the discrete label of the tower of field
excitations\foot{A good example is a $c=1$ model at the self-dual
radius, which may be interpreted as a string in a circle, or
as a string in a small $S^3$.}. For example, working in position
space \refs\rbranva, 
we could try to construct  SFT as standard  
  Field Theory  in the ``stringy" spacetime $ T^2_{(R_1,R_2)} \times 
T^2_{({\tilde R_1},{\tilde R_2})} = S^1_{R_1} \times S^1_{R_2} \times
S^1_{1/R_1} \times S^1_{1/R_2}$, since $K_0$ becomes simply a 
Laplace operator in the doubled space-time,  
\eqn\siete
{K_0 = -\left({\pt^2 \over \pt x_1^2} +
 {\pt^2 \over \pt{\tilde x}_1^2} 
+{\pt^2 \over \pt x_2^2} + {\pt^2 \over \pt {\tilde x}_2^2}\right) ,} 
where $(x_1, x_2)$ parametrize the torus $T^2_{(R_1,R_2)}$ and 
  $ ({\tilde x}_1, {\tilde x}_2)$ parametrize the T-dual torus. 
However, this is not quite correct,  because the new winding 
modes enter the level matching condition as 
\eqn\lm
{  (n_1\, \ell_1 + n_2\, \ell_2) + (L'_0 - {\overline L}'_0 )_L  
+(L_0 - {\overline L}_0 )_T + (L_0 - {\overline L}_0 )_{\rm 
ghost}  = 0.}
If we do not want to place additional constraints on the massless spectrum
of $D_T$ fields appearing in \dos\ in the decompactification limit, then
we may saturate the constraint within the zero mode sector:  
$n_1\,\ell_1 + n_2\,\ell_2 =0$. That is, we look at vertex operators
of the form 
$$
V_{\rm massless} = V_{1+1} \, e^{ip_L X_L} e^{ip_R X_R}
$$
where $V_{1+1}$ is a vertex operator for a massless state 
of the non-compact theory, satisfying \tres, and $p^i_{L(R)} = \half
\left( {n_i \over R_i} \pm \ell_i R_i \right)$. 
 In position space, we find that level matching
is realized at the massless level as an interesting ``chirality" 
condition:    
\eqn\lmm
{\left( {\pt^2 \over \pt x_1 \pt{\tilde x}_1} + {\pt^2 \over 
\pt x_2 \pt {\tilde x}_2} \right)\, \Psi_{\rm massless} = 0 .}
Notice that, it is only for those fields  saturating non-compact 
level matching that the condition on the longitudinal zero modes  
 takes this suggestive form. 
            In any case, it is clear that the determinant
of the operator $K_0$, in the space of functions satisfying the 
constraint \lmm, still fails to give the correct answer \uno.
A direct inspection of \uno\ reveals that we need to decouple the
field degrees of freedom in the torus $S^1_{R_1} \times S^1_{R_2}$
from those propagating in $S^1_{R_1} \times S^1_{1/R_2}$, with
T-duality acting only on the second circle. That is, we need to
enforce the decomposition:
\eqn\dec
{\Psi_{\rm massless} (x_1,x_2,{\tilde x}_1,{\tilde x}_2) = \Psi (x_1,
x_2) + \Psi (x_1, {\tilde x}_2 ).}
This does not follow from \lmm\ unless the level matching condition is
somehow further reduced to $\ell_2 n_2 =0$, by  
separately setting $\ell_1 =0$. Fortunately, there is a stringy  
mechanism for achieving just that. This is the subject of the next
subsection.

\subsec{Coset extensions}
On general grounds, a (perhaps infinite) set of free fields has a 
one loop partition function given by a determinant 
 of some kinetic  kernel $K$. We
can then write down a Schwinger representation of the form
\eqn\det
{{\rm log}\, Z_{\rm target} = -{1\over 2}\,{\rm Tr}' \,(-1)^F \, 
{\rm log}\, (K) =
{1\over 2}\, \int_{0}^{\infty} {ds\over s}\,{\rm Tr}' \,(-1)^F 
 \, e^{
-s\,K} }
conveniently regularized (for example, using a zeta-function
procedure), so that we 
obtain an ultraviolet
 and infrared finite quantity which we may compare with
the string computation. Indeed, eq. \det\ is reminiscent of a string
partition function 
\eqn\st
{{\rm log}\,Z_{\rm target} = {1\over 2} \, \int_{\CF} {d^2 \tau \over   
\tau_2^2} \, \CZ(\tau,{\bar\tau}) ,} 
where, essentially $\CZ(\tau,{\bar\tau})\sim 
 {\rm Tr} \, q^{L_0} {\bar q}^{{\overline L}_0}$,
and  
 the trace here may contain various GSO projections. There are,
however, two important differences between \det\ and \st. 
 The CFT  
partition function is integrated over  a fundamental domain 
$\CF$: $-{1\over 2} \le \tau_1 \le {1\over 2}$, $|\tau | \ge 1$ 
of the
genus one modular  group and, in particular, there is no
clear notion of ultraviolet region in such moduli space. The other  
difference is the presence of stringy states without particle analog,
such as  winding modes around compact dimensions, which nevertheless
we may try to incorporate either as extra fields in the spectrum of
$K$, as in the previous paragraph, or as additional ``small"
dimensions, as in \siete. Clearly, the first difference is
the most important one, because it implies that the  {\it  free}
SFT measure seems to be non-gaussian.  
 
The resolution of this puzzle is quite interesting. 
 By  a well known
mechanism \refs\rpol\rao, these two ``stringy" features tend to
cancel each other, at least partially. If the string theory is
sufficiently regular, we can trade the sum over {\it one} set
of winding modes by an extension of the integration region from the
fundamental domain $\CF$ to the strip $\CS$: $-{1\over 2} \le \tau_1 
\le {1\over 2}$, $\tau_2 \ge 0$. 
Technically, we just perform an extension of \st\ by cosets of the
translation group $\tau\rightarrow \tau +1$. In this way we 
 can tile the strip
$\CS$ with a set of $SL(2,{\bf Z})$ transforms of $\CF$, labeled
by two coprime integers $(c,d)=1$, of the form
 $\gamma_{(c,d)} ( \tau)
 = {a\tau +b \over
c\tau +d}$, where $a$ and $b$ are determined by the condition
$ad-bc=1$.
We then have $\CS= \cup \gamma_{(c,d)} (\CF)$, and the change
of variables in the integrand absorbs one of the winding sums,    
because a pair of integers can be replaced by a   
single  integer, together with   a pair of coprime integers. The basic
identity is the following
\eqn\coset
{\int_{\CF} {d^2 \tau \over \tau_2^2}\,F(\tau,{\bar\tau})\,
 \sum_{(\ell,\ell') \neq (0,0)} \,
2\pi R \, e^{-{\pi R^2 \over  \tau_2} |\ell' \tau  + \ell |^2 }    
= \int_{\CS} {d^2 \tau \over \tau_2^2} \, F(\tau,{\bar\tau})\,
\sum_{\ell\neq 0}\, 2\pi R \, e^{-{\pi R^2 \over \tau_2} \ell^2 }\,, }
where $F(\tau,{\bar\tau})$ is modular invariant.
The theta function on the left hand side of \coset\ is the classical
partition function of harmonic maps from the world-sheet torus to the
target circle $S^1_R$, while we find the particle analog on the right
hand side. Accordingly, all integers $\ell,\ell'$ represent winding
numbers. We can pass to a hamiltonian representation, based on the
mixed momentum-winding labels in \seis, by formally choosing a
``timelike" homology cycle on the torus, and applying a Poisson
resummation with respect to the corresponding winding numbers.

 The regularity
conditions ensuring the validity of this formula are essentially the
absence of infrared divergences driven by
 physical or unphysical  
tachyons, which would invalidate the exchange of sums and integrals
necessary to prove \coset\ (see \refs\rkutsei). 
The distinction between the two kinds of tachyons is of technical
nature.
Physical tachyons satisfy the non-compact level matching condition:
\eqn\nclm
{(L_0 -{\overline L}_0)'_L + (L_0 -{\overline L}_0)_T + (L_0 -{\overline
L}_0)_{\rm ghost} =0, } 
and they produce infrared instabilities in the non-compact theory.
On the other hand, unphysical tachyons are not present in the non-compact
theory either because they do not satisfy the non-compact condition \nclm, 
for example the tachyon in the bosonic sector in heterotic (0,$*$) models, 
or because they are projected out by GSO projections, as in the case of 
the NS tachyonic ground state in (1,$*$) heterotic strings.
However these unphysical tachyons may satisfy the
finite volume version \lm\ of the level matching condition with 
$n_1\ell_1+n_2\ell_2\neq 0$. The state resulting from this coupling
of the unphysical  tachyon with a winding-momentum state has a
$M_{\rm eff}^{2}$ which diverges in the decompactification limit but that
can vanish or even become negative at some finite values of the 
compactification radii, rendering the integral \st\ infrared divergent\foot{
The coupling of the NS tachyon with winding-momentum modes is ubiquitous in 
toroidal compactifications breaking space-time
supersymmetry.}. Whenever there are unphysical tachyons  
\st\ has to be defined with an  
an appropriate integration prescription, namely one
should perform the $\tau_1$ integral first. However, after the
transformation to the strip \coset, this prescription turns into a
complicated and not very useful
 integration rule over $\CS$. In practice, \coset\ holds only for
sufficiently large $R$, and singularities may develop at critical
values of the radius, when the winding unphysical tachyon
states develop a vanishing
effective mass $M^2_{\rm eff} = \sum_i (n_i^2 /R_i^2 + 
\ell_i^2 R_i^2 ) + M^2 =0$, with 
$M^2 $ given by \cuatro. A well known example of this phenomenon is
the standard Hagedorn thermodynamic  singularity, in which the corresponding
state becomes tachyonic in a whole region of the compactification moduli.
In other cases the state just becomes massless at one point and bounces
 back into
the $M^{2}_{\rm eff}>0$ region, producing singularities in the correlation
functions or its derivatives with respect to the moduli \refs\rosva.      
We will encounter examples of this latter type in section 3, 
and a more detailed  discussion can be found in  Appendix B.

Another condition on \coset\ is  the restriction to non-vanishing
winding number in the sums.   It corresponds to the subtraction
of the non-compact expression (the formal limit
$R \rightarrow\infty$ or, in other words, the non-compact
vacuum energy, with $R$ interpreted as euclidean time).
 This is important for the right hand side of
\coset\ to be well defined  in the  ($\tau_2 \sim 0$) ultraviolet  
region\foot{
The present subtraction is
rather natural in the context of string thermodynamics. It is well
known that  the non-vanishing vacuum energy of closed strings cannot
be written in field theoretical terms.  Of course,
this subtlety is trivial for 
supersymmetric strings with vanishing vacuum energy.}.

The important point is that the integrals over the strip are
reminiscent of \det, with $\tau_2$ being 
 proportional to the proper time
Schwinger parameter, and the integral over $\tau_1$ enforcing the
level matching condition. We then  conclude that, modulo appropriate
regularizations and vacuum subtractions, we can write down a standard
gaussian measure for free SFT, only 
 after  we have disentangled  precisely
 {\it one} set
of winding modes. Thus, there
 is a tension between  modular invariance and
gaussian measures, due to the ``temporal" winding states. This
mechanism provides an explanation of the doubling observed in \uno,
where only the relative T-duality between the two cycles (mirror
symmetry) is relevant. We now understand that, in order to write a
determinant (leading to the standard Dedekind function), we must use
up one set of winding modes (say around $S^1_{R_1}$) in going from
the $\CF$ representation to the $\CS$ representation.

\subsec{Doubling of vacuum partition functions}

In order to make these observations more precise, 
it is convenient to separate the ``transverse" CFT from 
the longitudinal one and various ghost 
systems, which cancel each other in the partition function (up to
discrete states of measure zero), and the contribution of the zero
modes.  We will suppose throughout this section that there is no
coupling between the transverse partition function and the
longitudinal zero modes. Such a coupling will appear in the next
section when we allow non-trivial boundary conditions of the target
fermions on the target torus. In what follows, we consider vacuum
(periodic) boundary conditions for all space-time fields. 
The resulting structure is ($q=e^{2\pi i\tau}$)  
\eqn\zetat
{\eqalign{{\rm log}&\,Z_{\rm target}
 = {1\over 2} \int_{\CF} {d^2 \tau \over
\tau^2_2} \, \Lambda_L (\tau,{\bar\tau}) \, \Lambda_{\rm ghost}
(\tau,{\bar\tau}) \, \Lambda_T (\tau,{\bar\tau}) \cr =& {1\over 2}
\int_{\CF} {d^2 \tau \over \tau_2^2 } \, \Lambda_T 
(\tau,{\bar\tau}) \,\tau_2 \,  \sum_{n_1,\ell_1}
 q^{{1\over 4} \left({n_1 \over R_1}
+\ell_1 R_1 \right)^2 }\, {\bar q}^{{1\over 4} \left( {n_1 \over R_1}
  - \ell_1 R_1 
\right)^2 }\,\sum_{n_2,\ell_2} q^{{1\over 4} \left( {n_2\over R_2} +
 \ell_2 R_2 
\right)^2 } \, {\bar q}^{{1\over 4} \left( {n_2\over R_2} - \ell_2
R_2 
\right)^2}     }}
 where $\Lambda_T (\tau, {\bar\tau})$ is the modular invariant non-compact
 transverse  partition function. It is constructed as a combination of
the  the holomorphic traces  
${\rm Tr}_{\CH_T}\, q^{L_0 - a}$ and ${\rm Tr}_{\CH_T} \, (-1)^F \, q^{L_0 -a}$, depending on the appropriate GSO projections\foot{The normal ordering
constant $a$ is the standard intercept, $a=1$ por a purely bosonic
sector, and $a=0,1/2$ in the Ramond and Neveu-Schwarz sectors
respectively.}, and similarly for the right moving CFT.  
     In particular, for the case of the $(2,*)$ strings, the
transverse partition function is purely holomorphic, and thus has the
general form  
\eqn\holo
{\Lambda_T (q)_{N=2} = D_T + C \, J(q),}
where $J(q)$ is the unique holomorphic modular invariant function with
a single pole at the origin and no zero mode, and $D_T$ is the effective 
number of
degrees of freedom. For example, for the $(2,0)$ string both constants
are non-trivial, whereas in the case of 
 the $(2,1)$ string leading to the
critical bosonic world-sheet, $C=0$ and $D_T =24$. For supersymmetric
target worldsheets, the non-compact 
transverse partition function vanishes by
supersymmetry. 
 
In order to use \coset\ we have to transform from the mixed
winding-momentum representation of \zetat\ to a complete winding
representation, by an appropriate Poisson resummation. 
The result, after going back to a mixed representation in the strip
is:
\eqn\ready
{{\rm log}\,Z_{\rm target} + C.T.  = {1\over 2}\,\sum_{n_1,n_2,\ell_2}\,
 \int_{0}^{\infty} {d\tau_2 \over
\tau_2} \left(\int_{-{1\over 2}}^{{1\over 2}} d\tau_1 \, 
\Lambda_T (\tau, {\bar\tau}) \, e^{2\pi i \tau_1 n_2 \ell_2} \right) \,
e^{-\pi\tau_2 \left( {n_1^2 \over R_1^2 } + {n_2^2 \over R_2^2 } +
 \ell_2^2 R_2^2
\right)} }  
where $C.T.$ stands for various counterterms, whose structure we will
detail below. 
In \ready, the  $\tau_1$ integral enforces
the level matching condition  \lm, without  the ``temporal" winding 
($\ell_1=0$), which has also dissapeared from the heat kernel. As a result,
any space-time interpretation involves a single circle of radius
$R_1$, in agreement with our comments above.
 In general, the partition function does not factorize in a natural
way . If we    want to make more precise statements, it is necessary
to further truncate the partition sum in \ready. For example, if we
restrict the loop trace to the states satisfying the non-compact level
matching condition \nclm,  
then we are effectively projecting
\eqn\repl
{\Lambda_T (\tau,{\bar\tau}) \rightarrow \int_{-{1\over
2}}^{1\over 2} d\tau_1 \, \Lambda_T (\tau, {\bar\tau}) 
. }
For $(2,*)$ strings, with holomorphic partition function, this
projection simply replaces $\Lambda_{T}(\tau)$ by 
the effective number of massless fields $D_T$. In more general cases we
are essentially inserting the ultralocal kernel
$$
\sum_f e^{-\pi\tau_2 \, M^2_f}\,,
$$
where the index $f$ labels the fields satisfying the non-compact level
matching constraint\foot{Notice that the projection \repl\ eliminates
off-shell tachyons from the partition sum.}. 

With these truncations, the effect of the $\tau_1$ integral in \ready\
is to set $n_2 \ell_2 =0$, which splits the partition function into
three terms:  
\eqn\two{\eqalign{
&{1\over 2} \sum_{f} 
 \int_0^{\infty} {d\tau_2 \over
\tau_2} \left[ 
\sum_{n_1,n_2}e^{-\pi\tau_2 \left({n_1^2 \over R_1^2 } +{n_2^2 \over
R_2^2} + M^2_f \right)}  \right. \cr
&+\left. \sum_{n_1,\ell_2}
e^{-\pi\tau_2 \left( {n_1^2 \over R_1^2} +
\ell_2^2 R_2^2 + M^2_f \right)} - \sum_{n_1}
e^{-\pi\tau_2 \left( {n_1^2 \over
R_1^2}
+ M^2_f \right)}  \right]\,. }}
In particular, if we perform a truncation to the massless $1+1$
fields, we finally obtain the desired result \uno: 
\eqn\fin
{\eqalign{{\rm log} Z_{\rm massless} =& -{D_T \over 2} \, {\rm
log}  \left[ \prod_{n_1,n_2}{}{'} \left( {n_1^2 \over R_1^2} +
{n_2^2 \over R_2^2} \right)  \prod_{n_1,\ell_2}{}{'} \left( {n_1^2
\over R_1^2} + \ell_2^2 R_2^2 \right) \prod_{n}{}{'} \left(
 {R_1^2 \over n^2 }
  \right) \right] \cr
=& -D_T \, \log {\left[ R_1\, |\eta (i{R_1/R_2})|^2 
\, |\eta(i R_1 R_2)|^2 \right]\,,}}} 
where we have discarded the zero mode terms in the sums in \two, and
defined the infinite products by zeta-function regularization. 
Note that the factorization exhibited in \fin\ is exact, in the
sense that it involves no truncations, when the partition function
is of the form 
\holo\ with $C=0$. This is the case of the $(2,1)$ models corresponding
to the bosonic and type II worldsheet in the vacuum (periodic) sector.
In the next section we show that the exact factorization is generally
lost when considering more general boundary conditions for the target
fermions.

We have been deliberately cavalier regarding regularizations 
in order to exhibit more clearly the important points. There are
however several subtractions of infrared and ultraviolet type involved. These
infinite counterterms relate our regularization and the ones implicit
in the calculation of \rdkl. 
         There is an infrared divergence already present in the
stringy expression \zetat,  when the string spectrum is purely
massless, coming from the trivial term in the zero mode sums. This
infinity is independent of the moduli $T,U$, and it is normaly
subtracted by the replacement\foot{ A somewhat disturbing feature of this
regularization is its lack of modular invariance.}
 $\Lambda_T (\tau,{\bar\tau}) \rightarrow
\Lambda_T (\tau,{\bar\tau}) -\tau_2 $ in \st. 
There are analogous logarithmic divergences, in this case picking both
infrared and ultraviolet
 components, in the expression to be integrated over the
strip in \ready. These terms are subtracted from \fin\
 in order to define the
determinants by the zeta function procedure:
$$
{\rm Tr}'\, {\rm log}\, K = -{d\over ds}\left. \int_0^{\infty} dt
\,t^{s-1} \,{\rm Tr}'\, e^{-t\,K} \right|_{s=0}\,.
$$
        Finally, there is a vacuum energy subtraction in \coset, in
the complete winding representation. Altogether, the complete
counterterm to be added to the zeta-function regularization is purely
ultraviolet and has the form
\eqn\cont{
C.T. = {1\over 2} \int_{\CS-\CF} {d^2 \tau \over \tau_2^2} \left(
R_1 \, f_{R_2} (\tau,{\bar\tau}) -\tau_2\right), }
where $f_{R_2}  (\tau, {\bar\tau}) $ denotes the integrand of the partition
function corresponding to the ``hamiltonian" target space ${\bf
R}\times S^1_{R_2}$.     The obvious asymmetry between $R_1$ and $R_2$
in the structure of \cont\ is related to our choice of $R_1$ as
``time", whose winding modes have been disentangled in going from
$\CF$ to $\CS$ using \coset.  The counterterm \cont\ ensures the
symmetry of the finite piece \fin.
                            In the supersymmetric
case, when $D_T =0$ and we have vacuum boundary conditions, $f_{R_2}
(\tau,{\bar \tau})$  is
just the vanishing vacuum energy, taking $R_1$ as the euclidean time
direction. In general, when the target theory has $D_T \neq 0$ or we
consider non periodic boundary conditions in the $S^1_{R_2}$ circle, 
as in the next section, then the
subtraction is non trivial.  
 A closely related quantity is the integral of the same function over
the fundamental domain $\CF$, which can be directly extracted from the
$R_1 \rightarrow \infty$ limit of \fin, with the result   
\eqn\casim
{\lim_{R_1 \rightarrow \infty}\, {1\over 2\pi R_1} \, 
{\rm log}\, Z_{\rm target} = D_T \, {1\over 12}  \, \left(
R_2 +{1\over R_2}\right).}
This has the appropriate form to be interpreted as a Casimir energy
contribution $-{\pi \over 6} {1\over 2\pi R_2}$ per degree of freedom
in the space $S^1_{R_2}$, plus the same contribution from $S^1_{1/R_2}$.  
For $D_T =1$ \casim\ coincides with the vacuum energy of the $c=1$
matrix model \refs\rbekleb.

\newsec{Target space spin structures} 

We have shown in the previous section that world-sheet
complexification in the sense of eq. \uno\ is an exact property of
$(2,*)$ models with $C=0$ in \holo, and an approximate property, in
the sense of the massless truncation \repl\ in all other cases.
This conclusion holds under the condition that no correlation exists
between the longitudinal zero mode labels and ``transverse" quantum
numbers is \zetat.
This means that all space-time fields have the same vacuum boundary
conditions as the bosons, namely periodic around both $R_1$ and $R_2$
target cycles. In what follows we examine the situation where
target fermions have antiperiodic boundary conditions around one or
both cycles. In other words, we want to obtain the torus spin
structures of the target space string. Because of fermion
antiperiodicity, we are led to consider finite temperature boundary
conditions in the underlying $(2,1)$ model. Obtaining in this way the
correct target space partition function as a function of the target
complex parameter $T$ is an interesting check of the stringy
construction of world-sheets. It also provides a testing ground of the
complexification phenomenon beyond the vacuum sector studied in the
previous section. We will find that, in general, the factorization of
$T$ and $U$ dependence is lost as an exact property, even in models
with a non-compact partitition function \holo\ with $C=0$. A
truncation to massless fields in the spirit of \repl\ is still
possible, but the procedure is somewhat ambiguous, and the
factorization depends on the particular projection.
 
In order to exhibit the details, let us consider the simplest
fermionic target string; the type-II model, which is constructed as
a $(2,1)$ background in terms of a left moving $\hat{c} =8$ SCFT with 
eight fermions together with eight bosons compactified in the root
lattice of $E_8$.  
In this theory, the target space statistics is completely determined
by the world-sheet parity of the $N=1$ left-moving sector. The
oscillator partition function is then given by \refs\rkutmar:
\eqn\dosbe{\Lambda_{s}^{s'} (\tau) = C(s',s)\,{E_4 (\tau)\over
 \eta^{12} (\tau)} \,
\,\half\,\vartheta^4 \left[\matrix{s' \cr s}\right] (0|\tau)\,,        
}
with $E_4 (\tau)$ the weight 4 Eisenstein series, which equals the
theta function for the $E_8$ root lattice \rkoblitz, and $C(s',s)$ the phases
giving the desired GSO projection ($C(\half,\half) =0$, $C(0,0) =
-C(0,\half) =-C(\half,0) = 1$). 

A modular invariant prescription for finite temperature boundary
conditions was introduced in ref. \refs\ratickw. Target fermions
become  antiperiodic
if the world-sheet spin structures $(s',s)$ are coupled to the winding
numbers $(\ell,\ell')$ around the corresponding ``thermal" circle by
the phases
\eqn\aw
{U^{s\,s'}_{\ell\,\ell'}=(-1)^{2s\ell+2s'\ell'+\ell \ell'}\,.}
These phases can be motivated as modular invariant extensions 
 of the standard phases in finite temperature Feymann
diagrams.
It is important to notice that this procedure only yields non-chiral
spin structures for the target fermions, namely both left and right
movers in the target are rendered antiperiodic by the phases \aw.
We shall postpone for the moment the discussion
 of chiral spin structures.

The general form of the partition function integrand \st\ is
\eqn\z
{
\CZ(\tau,\bar{\tau})=\sum_{\{s\}}\Lambda_{s}^{s'}(\tau)
\sum_{\{\ell\}} U^{s_{1}\,s'_{1}}_{\ell_{1}\,\ell'_{1}}\,
U^{s_{2}\,s'_{2}}_{\ell_{2}\,\ell'_{2}}\, 
e^{-S_{cl}(\ell_{1},\ell'_{1};\ell_{2},\ell'_{2})}=\sum_{\{s\}}\CZ^{cl}_{s,s'}
(\tau,\bar{\tau})
\,\Lambda_{s}^{s'}(\tau)
,}
where we have introduced a different set of phases for each target
circle. In this formula, when the cycle $S^1_{R_i}$ is periodic, we
set 
$U^{s_i s'_i}_{\ell_i \ell'_i} \rightarrow 1$ and $s_{j}=s$, $s_{j}'=s'$
for $j\neq i$. When both cycles are
antiperiodic (the (NS,NS) in the target), we set $s_1 =s_2
=s$, $s'_1 =s'_2 =s'$.

Finally, the classical 
action $S_{cl}$ is just the contribution of the classical
embeddings of the world-sheet torus onto the target torus with
winding numbers $(\ell_1,\ell'_1;\ell_2,\ell'_2)$
\eqn\clasa{S_{cl} = {\pi\over \tau_2} \,({\bar\tau}\ell_a + \ell'_a
)(g + b)_{ab}(\tau\ell_b + \ell'_b ).}
 When the target torus
is characterized by the complex moduli $T$ and $U$ we have
\eqn\scl{
\eqalign{e^{-S_{cl}}=&U_{2}\exp{\left[-{\pi U_2 \over T_2 \tau_2}\left[
|T|^{2}|\ell_1\tau+\ell'_1|^{2}+|\ell_2\tau+\ell'_2|^{2}+
2T_1 {\rm Re}(\ell_1\tau+\ell'_1)(\ell_2\bar{\tau}+\ell'_2)\right]\right]}
\cr 
\times & e^{2\pi iU_1 (\ell'_1 \ell_2 - \ell'_2 \ell_1)}\,}}
where $U_2$, the volume of the target torus, arises from the integration
of the zero modes in the path integral.

We have explained in  the previous section 
 how to go from the integral over
the fundamental domain $\CF$ to the integral over the strip $\CS$ by 
effectively setting one winding number to zero. This is also  going to work
 when the real parts of both $T$ and $U$ are switched on, as it is
the case now. So if we set $\ell_1=0$ in \scl\ and substitute in \z\
we  end up with the following expression for the classical part of 
the partition function in $\CS$
$$\eqalign{
\CZ^{cl}_{s,s'}(\tau,\bar{\tau})_{\CS}&=
U_{2}\sum_{\{\ell\}} \exp{\left[{-\pi U_2\over T_2\tau_2}[
|T|^{2} \ell{'}_{1}^{2}+|\ell_{2}'+\ell_{2}\tau|^{2}+2T_{1}\ell_{1}'(\ell_{2}'
+\ell_{2}\tau_{1})]+2\pi iU_{1}\ell_{1}'\ell_{2}\right]} \cr
&\times e^{2\pi i(\ell_{1}'s_{1}'+\ell_{2}s_{2}+\ell_{2}'s_{2}'+
{1\over 2}\ell_{2}\ell_{2}')}.}
$$
We can treat all the cases at the same time by remembering that
whenever the target space fermions have periodic boundary conditions 
along the $i$-th cycle ($i=1,2$) we have to set 
$s_{i},s'_{i}\rightarrow 0$ and $s'_{i}+{\ell_{i}\over 2}\rightarrow 0$,
while in the case (NS,NS)  we have $s_{i}=s$, $s'_i = s'$. 
After performing a Poisson resummation in $\ell'_{i}$ to go back to 
the mixed winding-momentum representation we find
\eqn\s{
\CZ^{cl}_{s,s'}(\tau,\bar{\tau})_{\CS}
=\tau_2\sum_{n_1,n_2,\ell_2}(-1)^{2 \ell_2 s_2}
q^{-\ell_2\left(n_2+s'_2+{1\over 2}\ell_2
\right)}
(q\bar{q})^{{1 \over 4 T_2 U_2}\left|n_1+s'_1-T\left(n_2+s'_2+{1\over 2}
\ell_2
\right)+U\ell_2\right|^{2}}}
in terms of which we can write $Z_{\rm target}$ as
$$
\log Z_{\rm target}={1\over 2}\int_{\CS}{d^{2}\tau\over \tau_{2}^{2}}
\sum_{\{s\}} {\cal Z}^{cl}_{s,s'}(\tau,\bar{\tau})_{\CS}
\,\Lambda^{s'}_{s}(\tau)
+{\rm counterterms.}
$$
The vacuum energy counterterm depends on whether the $S^1_{R_2}$ cycle
is supersymmetric or not. The density $R_1 \, f_{R_2} (\tau,
{\bar\tau})$ in \cont\ must be replaced by
$$
\sum_{\{s\}} \Lambda_s^{s'} (\tau) \sum_{\ell_2,\ell'_2} U_{\ell_2
\ell'_2}^{s_2 s'_2} \, e^{-S_{cl} (0,0;\ell_2,\ell'_2)}.           
$$
If the second cycle is periodic, then $U_{\ell_2 \ell'_2}^{s_2 s'_2}
\rightarrow 1$ and the vacuum subtraction vanishes because
of supersymmetry: $\sum_{\{s\}} \Lambda_s^{s'} (\tau) = 0$.

It will be useful to introduce the constants
\eqn\qs{
Q^{s'}_{s}=\int_{-{1\over 2}}^{1\over 2} d\tau_1\, \Lambda^{s'}_{s}(\tau)
,}
which measure the number of physical states in each spin structure 
for the non-compact theory; Jacobi's {\it aequatio} translates into the
identity
\eqn\aeq{
Q^{0}_{0}+Q^{1\over 2}_{0}+Q^{0}_{1\over 2}=0
.}
The  number of bosonic degress of freedom for the type-II 
model is then
just given by $-Q^{1\over 2}_{0} =8$.

In general, the resulting integral after $\tau_1$ integration is still
too complicated, and there is no obvious factorization of $T$ and $U$
dependence. We will nevertheless proceed as in the previous section,
defining  the massless truncation of the partition function by 
the projection  $\Lambda^{s'}_{s}\rightarrow Q^{s'}_{s}$ in \s, which
is analogous to \repl\ within each spin structure. Doing so
and integrating over $\tau_1$ leads to the analog of formula \two:  
\eqn\mass{\log Z_{\rm massless}={1\over 2}\int_{0}^
{\infty}
{d\tau_2\over \tau_{2}}\sum_{\{s\}}\sum_{n_{i},\ell_2}{}{'}\,
 Q^{s'}_{s}
\,(-1)^{2\ell_2s_2}\,\,(q\bar{q})^{A\bar{A}}
\,\,\delta_{\ell_2\left(n_2+s'_2+{\ell_2\over 2}\right),0}
}
where $A$ is given by
$$
A={1\over 2\sqrt{T_{2}U_{2}}}\left[n_{1}+s'_{1}-T\left(
n_{2}+s'_{2}+{\ell_{2}\over 2}\right)+U\ell_{2}\right].
$$
The remaining integral can be reduced to a sum of logarithms of infinite
products which can be computed using zeta-function regularization
(the details of the computation are worked out in Appendix A). The
final result in the four sectors can be written as
\eqn\res{\eqalign{Z_{\rm (R,R)}&=0 \times 1 \cr
Z_{\rm (NS,R)}&=\left[f_{2}(T)f_{2}(U)\right]^{4} \cr
Z_{\rm (R,NS)}&=\left[f_{4}(T)f_{4}(2U)\right]^{4} \cr
Z_{\rm (NS,NS)}&=(q_{T}q_{2U})^{-{1\over 8}}
\left[f_{3}(T)f_{3}(2U)\right]^{
4},}}
where
$f_i (T)$ is given in terms of standard Jacobi theta functions:
$$
f_{i}(T)\equiv {\theta_{i}(0|T) \over \sqrt{T_2}\eta^{3}(T)}
.$$ 
We have extracted the  holomorphic square root of $Z_{\rm massless}$ in
\res\ in order to exhibit only the chiral structures. The Ramond
sector zero mode has been included by hand in the first line of \res,
 since  
the $(2,1)$ string torus diagram  only sums up the non-zero modes in
target space,
whose overall contribution is one, due to the supersymmetric
boson-fermion cancellation, i.e. $D_{T}=0$.

It is interesting that we get the correct $T$-dependent terms; the
modular invariant combinations of Jacobi and Dedekind functions with
the right multiplicity to be interpreted as building blocks of the
target type-II worldsheet\foot{Notice, however, the occurrence of a
target modular anomaly in the (NS,NS) sector, $(q_T q_{2U})^{-1/8}$,
whose $T$-dependent term does not disappear in the field theory limit
$\alpha' \rightarrow 0$.}. The method used to introduce antiperiodic
boundary conditions in the target fermions always produces left-right
symmetric structures in the target, i.e. $|f_i (T)|^2$. A fully
stringy construction of more general asymmetric terms like $f_i (T)
\overline{f_j (T)}$ is a more complicated issue on which we offer
some comments in the next subsection. 
 In general, additive structures in $Z_{\rm
target}$ are beyond the reach of $(2,1)$ string perturbation theory,
such as the rules for combining the different target spin structures
(target GSO projection) or the global structure of the target bosons
(effects due to compactness of the target bosons are of order
$e^{-1/\alpha'_{\rm target}} \sim e^{-1/\lambda^2}$).

  Regarding the $U$ dependence,
 \res\ has a second         
piece that only in the (R,R) and (NS,R) sectors can be interpreted as the 
corresponding partition function in the mirror torus. There
is a  simple rule to get these $U$-terms; in the sectors ($*$,R) 
they are obtained from the $T$-terms by a T-duality transformation
in the second cycle, 
$T\rightarrow U$. For the ($*$,NS) cases the second 
term is obtained instead by $\beta$-duality, $T \rightarrow 2U$ 
(see Appendix B). 
In addition, it seems that
 whenever we have different boundary conditions in the
two cycles,  the result depends on what cycle we choose to
disentangle the windings to pass from the fundamental domain
to the strip. The condition for the choice to be irrelevant is 
that 
$f_{2}(U)$ should 
be invariant under the Atkin-Lehner transformation \refs\rmal,  
$
f_{2}(U)=f_{2}(-1/ 2U)
$,  
condition which is not fulfilled in our case.  As a result,  the 
$U$-dependence seems to be  
 sensitive to the order in which we truncate the 
theory. 

These results may seem rather puzzling at first sight. It is not difficult
to convince oneself that the $\CF$-representation of the target partition 
function \z\ is invariant under 
T-duality in the cycles with R boundary conditions and $\beta$-duality
in those with NS. On the other hand, our expression
\res\ only has one of those invariances, while the one associated with the
cycle used to disentangle the winding modes is broken\foot{Incidentally,
the result \res\ in the (NS,NS) sector is manifestly $\beta$-duality
invariant in both cycles except for the {\it anomalous} term $(q_T
q_{2U})^{-1/8}$, which should be then interpreted as an artifact of the
truncation.}. The origin of 
this breaking can actually be traced back to the 
projection $\Lambda^{s'}_{s}\rightarrow Q^{s'}_{s}$ made in order to compute 
the integral over the strip.
This projection retains only the left-right symmetric states of the
non-compact theory in the transverse partition function, and the final
result is sensitive to the asymmetry between $R_1$ and $R_2$ cycles in
the mixed spin structures.  
On the other hand, if we were to compute the {\it full} integral
over $\CF$ without any projection we would find  expressions 
diverging logarithmically as 
the radius of any cycle with antiperiodic boundary conditions
approaches  
the self-dual value under $\beta$-duality. Only the contribution 
of the (R,R) spin structure  vanishes identically even before integration 
and therefore will be regular for all value of the moduli. The origin 
of these divergences lies in the existence of some winding state  
becoming massless at special values of the moduli (the off-shell
tachyons mentioned in section 2). As we discuss in  more
detail in Appendix B, a careful analysis of the $\CF$ and the projected
$\CS$ representation together with the type of singularities encountered in
the modular invariant expression, implies that the result of performing
the {\it full} integral over $\CF$ of the partition function \z\ does not
factorize into two parts depending only on $T$ and $U$ respectively. 
We would obtain  the factorized contribution \res\ plus
a collection of non-factorizable terms $F(T,U)$ which diverge logarithmically
on some codimension one submanifolds of the target moduli space. 

The situation is reminiscent of the one in (2,0) models. The computation 
of $\log Z_{\rm target}$ in ref. \refs\rdkl\ gives,
in addition to the contribution \uno\ coming from the $1+1$  massless 
states, a term
$-\log |J(T)-J(U)|^{2}$ which spoils factorization of the integrated 
partition function. This latter term represents the 
contribution of  states which do not satisfy the non-compact level
matching condition \nclm. Here we also find a logarithmic divergence
triggered by winding states becoming massless whenever one of the radii
reaches the self-dual value under T-duality, $R_{i}=1$ \refs\rosva.

\subsec{Chiral spin structures and the target heterotic string}

In this section we present some observations on the problem of
constructing fully chiral spin structures for the target string.
As we already pointed out previously,
the standard finite-temperature phases \aw\ render both left and right
moving target fermions antiperiodic around a given cycle.
The problem of handling chiral spin structures is even more relevant 
in the case of the $(2,1)$ model which reproduces the world-sheet theory
of the $(1,0)$ heterotic string. In this case both matter and gauge 
fermions have only one chirality, so there is no {\it symmetric} sector 
which could be obtained by using the phases $U^{ss'}_{\ell\ell'}$.  

The correct construction must treat the zero modes chirally in the
target space sense, and it 
 is natural to expect that a more sophisticated asymmetric orbifold
(cf. \refs\rnsv) is needed in this case. We can try to guess some
pieces of the full answer from the structure of eq. \mass. Focusing on
the field theory limit for simplicity, we set $\ell_2 \rightarrow 0$
and concentrate on the $T$-dependence (all $U$-dependence is reduced to an 
additive term $\log{U_{2}}$). Then, the generalization of \mass\
leading to chiral target space structures has the form  
\eqn\lz{
\log Z_{\rm massless}={1\over2}\int_{0}^{\infty} {d\tau_{2}\over \tau_{2}}
\sum_{\{s\}}\sum_{n_{i}}{}'\,\,Q^{s'}_{s}\,\,(q\bar{q})^{A_{L}A_{R}}
}
where now  $A_{L}$ and $A_{R}$ are not in general complex conjugate of
each other, but depend on the boundary conditions for the target left and 
right moving fermions,  
\eqn\pres{
\eqalign{
A_{L}&={1\over 2\sqrt{T_{2}U_{2}}}\left[n_{1}+s_{1}'-T(n_{2}+s_{2}')\right] 
\cr
A_{R}&={1\over 2\sqrt{T_{2}U_{2}}}\left[n_{1}+t_{1}'-
{\overline T}(n_{2}+t_{2}')\right]
.}
}
Now $s_i,(t_i)\rightarrow 0$ whenever the left (right) moving fermions are
periodic along the $i$-th cycle whereas in the case in which the
left (right) moving fermions are antiperiodic in both cycles we will
have $s_{i}=s$, $s_{i}'=s'$ ($t_{i}=s$, $t_{i}'=s'$). It is easy to 
check, using the formulae of Appendix A, that the computation of $\log Z_{
\rm massless}$ gives the correct chiral terms $f_{i}(T)\overline{f_{j}(T)}$.

The situation is much more complicated if we retain the dependence on 
$U$, or in other words,  if we work at a finite value of $\alpha'$. 
The reason
being that when $\ell_{2}\neq 0$,
 \mass\ contains a phase $(-1)^{2\ell_{2}s_{2}}$
and therefore there is an ambiguity in how to split this phase into 
contributions from left and right moving target fermions.  
 A related
question is how to promote the delta function imposing the level
matching in \mass\ to a fully modular invariant integrand.

Another aspect of the subtleties with chiral target fields arises when
considering the oscillator partition function of the $(2,1)$ model
leading to the target heterotic string.  
In \refs\rkutmar \refs\rkutmaro\  this vacuum  
was constructed as an orbifold of the  type-IIB
background. The world-sheet degress of freedom
of the (2,1) string\foot{Our notation
here follows that of \refs\rkutmaro. $x^{0},\ldots,x^{3}$ are
the coordinates in the dim=$2_{\bf C}$ space-time with signature
$(--++)$ and $y^{i}$ are the coordinates in the internal left moving
$E_{8}$ torus. In addition we have their fermionic partners $\psi^{0},\ldots,
\psi^{3}$ and $\lambda^{i}$.} are the 
complex fields  $Z^{\mu}=x^{\mu}+ix^{\mu+1}$, $\Psi^{\mu}=
\psi^{\mu}+i\psi^{\mu+1}$ ($\mu=0,2$) and their corresponding right moving 
components, together with the $N=1$ left moving superfields $\Phi^{i}=
y^{i}+\theta\lambda^{i}$ ($i=1,\ldots,8$) in the internal $\hat{c}=8$
SCFT. Let us compactify all open dimensions except $x^1$.
The orbifold construction leading to the target $(1,0)$ heterotic
string is defined by (see  \refs\rkutmaro)
\eqn\or{
\eqalign{Z^{\mu}&\longrightarrow Z^{\mu\,*} \cr
\Psi^{\mu}&\longrightarrow \Psi^{\mu\,*} \cr
\Phi^{i}&\longrightarrow -\Phi^{i}
}
}
and the analog for $\overline{Z}^{\mu}$, $\overline{\Psi}^{\mu}$. If we 
gauge the left moving current $J=\partial x^{1}+\partial x^{3}$ we get
the fermionic version of the $SO(32)$ heterotic string, whereas if we take
$J=\partial x^{1}+\partial y^{1}$ what we find is the $E_{8}\times E_{8}$ 
heterotic string at finite coupling.

We now compute the partition function for the orbifold associated with 
the $SO(32)$ heterotic string. In the 
$N=2$ right moving sector this was done by Mathur and Mukhi in
\refs\rmm. What they found is that in each sector of boundary conditions
the contribution from the matter fields is exactly canceled by the
ghosts, giving a constant partition function. The computation on the 
right moving $N=1$ part is more subtle because of the gauging of the null 
current $J$. Here we have, in addition to the $N=1$ ghost system $(b,c)$ and
$(\beta,\gamma)$, a (1,0) ghost associated with the $U(1)$ gauge current
$(t,\bar{t})$ and its $N=1$ superpartner of spin (1/2,1/2), $(f,\bar{f})$.
Altogether, we have effectively two light-cones, and four sets of
oscillators are canceled out. The zero modes in the $(x^1, x^3)$ plane
are also eliminated against the integration over $U(1)$ Wilson lines,
and we are  essentially left 
only with the transverse $\hat{c}=8$ 
internal excitations. A somewhat delicate point is  
the determination the statistics
of the target space fields in the space-time $(x^{0},x^{2})$. From the
analysis of reference \refs\rmm, it can be seen that in the untwisted
sector the statistics is determined solely by the internal $\hat{c}=8$ SCFT,
whereas in the twisted sector the ground state in the $N=2$ part is 
fermionic and therefore one has to take into account both left and right
movers. After a rather simple computation one arrives at the following
 result:  
\eqn\orb{
\eqalign{\CZ_{\rm SO(32)}&=\left[{\theta^{4}_{3}-\theta_{4}^{4} \over 
2\eta^{12}}\left({E_{4}+\theta_{3}^{4}\theta_{4}^{4}\over 2}\right)-
{\theta_{2}^{4}\over 2\eta^{12}}\left({E_{4}-\theta_{3}^{4}\theta_{4}^{4}
\over 2}\right)
\right.\cr 
&\left.-{\theta^{4}_{3}+\theta_{4}^{4} \over \eta^{12}}\left({\theta_{3}^{4}
\theta_{2}^{4}+\theta_{4}^{4}\theta_{2}^{4}\over 2}\right)+
{\theta_{2}^{4}\over
\eta^{12}}\left({\theta_{3}^{4}
\theta_{2}^{4}-\theta_{4}^{4}\theta_{2}^{4}\over 2}\right)\right]
.}}
The first two terms in the right-hand side correspond to the contribution
from the untwisted sector while the last two come from
the twisted sector. There is a relative factor of $2$ in the twisted
sector,  
from the total number of fixed points
 in the non-chiral part of the orbifold. 
The twisted internal  bosons $y^i$ 
 yield inverse powers of Jacobi theta functions,
which can be transformed to the numerator using the identity $\theta_2
(\tau)\theta_3 (\tau) \theta_4 (\tau) = 2\,\eta (\tau)^3$. The
corresponding factors of $2$ cancel the chiral fixed point degeneracy
$(\sqrt{ 2^8})$.
Finally, we should multiply \orb\ by the solitonic partition function
associated with the $(x^0,x^2)$ torus,  $Z_{2,2}(T,U)$. 

Although $Z_{2,2} (T,U)$ is apparently inert under \or,
 the structure of  the massless states in \orb\  suggests that
the chiral structure in target space has to be imposed on $Z_{2,2}
(T,U)$ by hand, correlated with the twisting in \or. In 
the twisted sector it is easy to check that we have
 only 32 fermionic states,  
which correspond to the gauge fermions of the $SO(32)$ heterotic string. 
In fact, using some elementary properties
 of the Jacobi theta functions,  
it is possible to show that the whole twisted sector adds to a 
constant\foot{It is interesting to note that a different sign
convention is possible in the twisted sector, compatible with modular
invariance, giving an overall positive constant which would be
interpreted as the number of chiral bosons in the bosonic formulation
of the heterotic world-sheet. For this to work, we have to consider
non-compact $(x^1, x^3)$ directions, resulting in only one orbifold
fixed point. This removes the relative factor of 2 in \orb, thus
producing a constant twisted partition function equal to $+16$.} 
equal to $-32$. In the untwisted sector, however, the interpretation is
not so straightforward. We find 8 bosons as one would
expect (the standard $\theta_3^4 -\theta_4^4$ term),
 but  there are no
massless fermions (the term proportional to $\theta_2^4$ has no zero
mode). This seems to be in conflict with the analysis of 
\refs\rkutmaro\ in which it is found that after projecting onto the states 
 invariant under the modding \or, we are left with 8 fermions with 
a definite chirality.
A solution to this apparent paradox would be  that the oscillator part
of the  partition
function \orb\ is not sensitive  to the target
 space chirality of the fields, which should in turn be associated
with selection rules in the $Z_{2,2} (T,U)$ solitonic terms. From this point 
of view,  projecting out  
8 chiral fermions in the untwisted sector results in the elimination of
8 non-chiral fermions from the partition function, the number we began with.
This would be resolved by adjoining different solitonic factors to the
two terms $E_4$ and $\theta_3^4 \theta_4^4$, thereby correlating the
modding \or\ in the directions $(x^1, x^3, y^i)$ with a
``longitudinal" asymmetric orbifold in the $(x^0, x^2) $ target
world-sheet. The question is to find a
modular invariant procedure to do this.  

With this in mind, we may put together a set of {\it ad hoc} rules,
based on \pres, to obtain the chiral terms in the target partition
function.  
As we pointed out before, any determination of the relative signs between
spin structures is beyond the perturbative analysis of the (2,1) heterotic
string and can only be addressed by making use of non-perturbative information.

\newsec{Concluding remarks}                                                   

In the present paper we have studied some of the properties of the world-sheet
theories emerging from the target space dynamics of (2,1) heterotic strings.
In particular we have been specially concerned with the apparent 
doubling of the target world-sheet degrees of freedom. We have described
the physical mechanism underlying this doubling in the vacuum sector of
the partition function, and found  it to be a rather generic feature
of certain massless truncations of two-dimensional 
 strings. The factorization is exact for the vacuum sector of the
$(2,1)$ models leading to the bosonic and type-II world-sheets.
In the case of the (2,0) string such doubling
happens only after performing the projection \repl\ onto the states that 
satisfy the non-compact level matching condition \nclm. 
We have also considered non-trivial boundary conditions for the target
fermions of the target type-II model. In general, the $T$-$U$
factorization does not hold 
 in this sector, whose behavior is similar to the
$(2,0)$ string. There are several ``natural" massless projections
which produce a $T$-$U$ factorized result, although it seems that the
universality of the vacuum sector is lost.
A partial discussion of the heterotic target string is offered,
including the calculation of the oscillator partition function of the
corresponding $(2,1)$ asymmetric orbifold, and some remarks on the
handling of chirality in target space.

Despite its limitations, the target world-sheet doubling found in the
bosonic sector is rather significative. It 
 is interesting to elaborate on its  geometric interpretation. 
 According to our target space
analysis in section 2, the level matching condition is satisfied
by propagating fields in the space   
$S^1_{R_1} \times S^1_{R_2} \times S^1_{1/R_2}$ 
with the additional chirality constraint,
\eqn\qui{
{\pt^2 \over \pt x_2 \pt {\tilde x_2}} \Psi_{\rm massless} =0         
.}                                                                    
So, we can only double {\it space}, and we still have to impose a  
chirality condition on the field excitations in this ``stringy"
space-time. 
Equation \qui\ is now solved by fields  living in factorized
Hilbert spaces
\eqn\hil{
\CH = \CH (S^1_{R_2}) \otimes \CH (S^1_{1/R_2})
.}
This interpretation is also suggested by the Casimir energy in the
``hamiltonian'' manifold ${\bf R} \times S^1_{R_2}$, eq. \casim. 
Because of conformal invariance in the target, the Hamiltonian
evolution  depends only
on the combinations $T_2 = R_1 /R_2$ and $U_2 = R_1 R_2$. If we switch
 on
the torsion, 
the 
complete ``twisted" euclidean evolution operator on a cylindrical
target space becomes  
\eqn\compl{
(q_T q_U)^{L_0^{\rm target} - D_T /24} \,\,\, 
({\bar q}_T \, {\bar q}_U
)^{{\overline L}_0^{\rm target} - D_T /24}
.} 
We may use these operators as building blocks for an operator 
formalism sewing higher order target world-sheets, and also to insert
 external scattering states. We  
conclude that local operator insertions also have  complexified moduli.
Accordingly, the vertex operators of the target string must be 
integrated over the world-sheet moduli $z_i$ and the complex
conjugated ${\bar z}_i$
independently. This is exactly the structure emerging from the
analysis of generalized high-energy scattering saddle points, as 
explained in \refs\rwit!.   Following \refs\rgrossm, the high energy
asymptotics of string scattering is dominated by saddle points in the
complete world-sheet moduli space. In particular, there is a
``dominant"
location of the vertex operators on the world-sheet. It was pointed
out  by
Witten in \refs\rwit\ that, by allowing both  momentum and winding
modes for the external particles, the most general saddle point
configuration is obtained by varying independently the holomorphic
positions of the vertex operators.

It is then natural to suspect that  perhaps the $(2,1)$ construction
gives directly the   high energy
phase of standard strings. It would be very interesting to make this
correspondence more explicit. Notice that the high energy behavior is
only determined by the bosonic sector of the colliding strings,
because the saddle points only depend on the tachyonic part of the
vertex operators $e^{ipX}$. Perhaps this explains why we find
difficulties in implementing the complexification beyond the bosonic
sector of the target space strings.
A potentially interesting observation is the following: formally, the
high energy limit of the target string world-sheet dynamics
corresponds to the limit $\alpha'_{\rm target} \sim \alpha' \lambda^2
\rightarrow \infty$, (see \refs\rgrossm), which naturally singles out
the strong coupling limit of the $(2,1)$ model.

If we take the complexification idea seriously, we should 
 understand how these  results follow from a gauge
fixing procedure of target space gravity. The required doubling of
degrees of freedom has to do with the existence of a
graviton-axion-dilaton system both in momentum variables,  
$
\xi_{ab} \pt X^a {\bar \pt} X^b e^{ipX}
$
and in winding variables, ${\tilde \xi}_{ab} \pt {\tilde X}^a {\bar
\pt}{\tilde X}^b e^{i{\tilde p}{\tilde X}}$ .
 In a weak field expansion, our previous remarks
 indicate that we only have one independent ``winding"
gravitational system, because the windings around one of the 
cycles have been used to write the target space dynamics as field
theory dynamics.
It would be interesting to make these remarks more precise. In
addition, the complexification in \compl\ should be tested at higher
orders in the target topological expansion.

An interesting problem that we have not dealt with here is the 
three-dimensional generalization
(membrane world-volume). The complexification phenomenon in the sense
of \compl\ is essentially two-dimensional. The reason being that the
Hilbert space decoupling \hil\ only occurs in this case. For a higher
dimensional toroidal compactification, the level matching condition
saturated over zero modes becomes $$ \sum_{i=2}^{d} {\pt^2 
 \over \pt x_i
\pt {\tilde x}_i} \Psi_{\rm massless} =0, $$ 
where we have already disentangled the ``temporal" winding modes
$\ell_1 =0$. So, there is a certain chirality condition imposed
over the spatial dependence of the fields, but the geometric
interpretation is certainly more complicated than the two dimensional
case.

\newsec{Aknowledgements}
 
It is a pleasure to thank J. Distler, D.J. Gross, I.R. Klebanov, 
S. Mukhi, M.A.R. Osorio, S. Ramgoolam and H. Verlinde for useful
discussions and comments. The  work of J.L.F.B. 
 was supported by NSF PHY90-21984 grant.
The work of M.A.V.-M. was supported in part by a Spanish MEC
postdoctoral fellowship.

\appendix{A}{Computation of determinants}

In this appendix we will detail the computation of the integral \mass. The 
first important thing to notice is that the phase $(-1)^{2\ell_2s_2}$ is
trivial; the Kronecker delta function enforces either $\ell_2=0$ or
$\ell_2=2(n_2+s'_2)$,  so that $2\ell_2s_2=4s_2s'_2$ mod 2,  which can
only be odd if $s_2=s'_2={1\over 2}$. But this corresponds to the 
contribution of the odd spin structure which vanishes ($Q^{1\over 2}_{1\over 2}
=0$). Then we can forget about this phase in the subsequent computation.

The integral \mass\ can then be easily written in terms of infinite products
as
\eqn\prodc{\eqalign{ \log Z_{\rm massless} &= -{1\over 2} \sum_{s,s'}
Q_{s}^{s'} \log\left[\prod_{n_1,n_2}{}^{'}{|n_1+s'_1-T(n_2+s'_2)|^{2}\over
T_2U_2} \prod_{n_1,\ell_2^{c}}{}^{'}{|n_1+s'_1-U\ell_2^{c}|^{2}\over
T_2U_2}\right] \cr
&+{1\over 2}\sum_{s,s'}Q^{s'}_{s} \delta_{s'_2,0}\log
\prod_{n_1}{}^{'}{(n_1+s'_1)^{2}\over T_2U_2} + {\rm constants}}
}
where
$$
\ell_{2}^{c}=\left\{\matrix{\ell_2 & \hbox{in the ($*$,R) case} \cr
2(n_2+s'_2) & \hbox{in the ($*$,NS) case}}\right.
$$
and the primes in the products indicate that we omit the zero mode
whenever there is one.

In order to compute the previous expression using zeta-function regularization
we start with the well-known identities
$$\eqalign{
\prod_{n=1}^{\infty}\left(A+{n^{2}\over B}\right)=&{2\over\sqrt{A}}\sinh
\pi\sqrt{AB} \cr
\prod_{n=0}^{\infty}\left(A+{\left(n+{1\over 2}\right)^{2}\over B}\right)
=&2\cosh\pi\sqrt{AB}
}$$
and using them we write the relevant infinite products in terms 
of standard theta-functions
$$\eqalign{ \prod_{m,n}{}^{'} |n-mT|^{2} &= \left|{\vartheta'\left[
\matrix{1/2 \cr 1/2}\right](0|T) \over \eta(T)}\right|^{2} =
 4\pi^{2}|\eta(T)|^{4} \cr
\prod_{m,n}\left|n+a-\left(m+b\right)T\right|^{2}&=
(q_{T}\bar{q}_{T})^{-{1\over 8}ab}
\left|{\vartheta\left[\matrix{{b+1/2} \cr {a+1/2}}\right](0|T) 
\over \eta(T)}\right|^{2},
}
$$
where $a,b=0,{1\over 2}$ but not simultaneously zero.

Now we can proceed with the computation in the four different sectors.
The easiest one is the (R,R)
$$\eqalign{
\log|Z_{\rm (R,R)}|^{2}&=-{1\over 2}\left(\sum_{s,s'}Q^{s'}_{s}\right)
\log\left[\prod_{m,n}{}^{'} {|n-mT|^{2}\over T_2U_2}
\prod_{m,n}{}^{'} {|n+mU|^{2}\over T_2U_2}
\prod_{n}{}^{'} {T_2U_2 \over n^{2}}\right] \cr
&=-\left(\sum_{s,s'}Q^{s'}_{s}\right)\log\left[
\sqrt{T_{2}}|\eta(T)|^{2}\sqrt{U_{2}}|\eta(U)|^{2}\right]=0
}
$$
because of \aeq . The other cases are less trivial; for example, in the
(NS,R) and (R,NS) sectors we find using again \aeq\
$$\eqalign{
\log|Z_{\rm (NS,R)}|^{2}&={8\over 2}\log\left[{1 \over T_{2}U_{2}}{
\prod_{m,n}\left|n+{1\over 2}-mT\right|^{2} \over 
\prod_{m,n}^{'}\left|n-mT\right|^{2}} 
{\prod_{m,n}\left|n+{1\over 2}+mU\right|^{2} \over 
\prod_{m,n}^{'}\left|n+mU\right|^{2}}\right] \cr
&=4\,\log\left[\left|{\theta_{2}(0|T)\over \sqrt{T_{2}}\eta^{3}(T)}
\right|^{2}\left|{\theta_{2}(0|U)\over \sqrt{U_{2}}\eta^{3}(U)}
\right|^{2}\right]
}
$$
and
$$\eqalign{
\log|Z_{\rm (R,NS)}|^{2}&={8 \over 2}\log\left[{|T|^{2}|2U|^{2}
\over 2T_{2}U_{2}}{
\prod_{m,n}\left|m+{1\over 2}-n\left(-{1\over T}\right)\right|^{2} \over 
\prod_{m,n}^{'}\left|m-n\left(-{1\over T}\right)\right|^{2}} 
{\prod_{m,n}\left|m+{1\over 2}+n\left(-{1\over 2U}\right)\right|^{2} \over 
\prod_{m,n}^{'}\left|m+n\left(-{1\over 2U}\right)\right|^{2}}\right] \cr
&=4\,\log\left[\left|{|T|\theta_{2}\left(0\left|
-{1\over T}\right.\right)\over \sqrt{T_{2}}\eta^{3}\left(-{1\over T}\right)}
\right|^{2}\left|{|2U|\theta_{2}\left(0\left|
-{1\over 2U}\right.\right)\over \sqrt{2U_{2}}\eta^{3}\left(-{1\over 2U}\right)}
\right|^{2}\right] \cr
&=4\,\log\left[\left|{\theta_{4}(0|T)\over \sqrt{T_{2}}\eta^{3}(T)}
\right|^{2}\left|{\theta_{4}(0|2U)\over \sqrt{2U_{2}}\eta^{3}(2U)}
\right|^{2}\right].
}
$$
Finally we have to go to the (NS,NS) sector. The procedure is exactly
the same and the final result is
$$
\eqalign{
\log|Z_{\rm (NS,NS)}|^{2}&={8 \over 2}\log\left[
{
\prod_{m,n}\left|m+{1\over 2}-\left(n+{1\over 2}\right)T\right|^{2} \over 
2T_2 U_2 \prod_{m,n}^{'}\left|m-nT\right|^{2}} 
{\prod_{m,n}\left|m+{1\over 2}+\left(n+{1\over 2}\right)(2U)\right|^{2} 
\over 
\prod_{m,n}^{'}\left|m+n(2U)\right|^{2}}\right] \cr
&=4\,\log\left[\left|q_{T}^{-{1\over 32}}{\theta_{3}(0|
T)\over \sqrt{T_{2}}\eta^{3}(T)}
\right|^{2}\left|q_{2U}^{-{1\over 32}}{\theta_{3}(0|
2U)\over \sqrt{2U_{2}}\eta^{3}(2U)}
\right|^{2}\right].
}
$$
All these formulae hold up to finite additive numerical constants.

\appendix{B}{$\beta$-duality vs. T-duality}

T-duality is a well known symmetry of the toroidal compactifications of 
both bosonic and
heterotic strings. The theory is symmetric under the following combined
transformation of the compactification radius and the string coupling constant
\refs\rtdual
\eqn\tt{
R \rightarrow {\alpha' \over R} \hskip 1cm \lambda \rightarrow {\sqrt{\alpha'}
\over R}\, \lambda 
.}
In the case of the type-II superstring the previous transformation is
not a symmetry of the theory, but rather maps the type-IIA into the type-IIB
and viceversa \refs\rseipol. 

In the case of fermionic strings one can think of more general toroidal 
compactifications in which the boundary conditions of the target space 
fermions are correlated with those of the world-sheet fermions. A particular
class of such compactifications is that in which space-time fermions are
taken to be antiperiodic along the compactified dimension \refs\rramp\ratickw.
These boundary conditions are implemented in the correlation functions
by the insertion of a set of phases in the sum over spin structures
and over classical vacua of the two-dimensional world-sheet theory. At genus
$g$ these phases are \refs\ratickw
$$
U_{\vec{\ell},\vec{\ell}^{\,'}}^{\vec{s},\vec{s}^{\,'}}= 
(-1)^{2\vec{\ell}\cdot\vec{s}
+2\vec{\ell}^{\,'}\cdot\vec{s}^{\,'}+\vec{\ell}\cdot\vec{\ell}^{\,'}}
$$
where $\vec{\ell}$, $\vec{\ell}^{\,'}$ are the winding numbers 
of the $2g$ homology cycles $\{a_{i},b_{i}\}$ ($i=1,\ldots,g$)
into the target circle $S^{1}_{R}$ and $\{\vec{s},\vec{s}^{\,'}\}$
are the corresponding spin structures. 
In the case of a heterotic string theory, in which space-time
fermions arise only from the left moving sector,
 we have  to introduce 
only one set of phases. Then  
the genus-$g$ contribution to the logarithm of the partition function can 
be written as
\eqn\zz{\eqalign{
[\log Z(R)]_{g}&=\lambda^{2g-2}\int_{\CF_{g}}
d\mu(\tau)\sum_{\vec{s},\vec{s}^{\,'}}
\Lambda_{\vec{s}}^{\vec{s}^{\,'}}(\tau,\bar{\tau}) \cr &\times
\sum_{\ell_{i},\ell_{i}^{'}}(-1)^{2\vec{\ell}\cdot\vec{s} 
+2\vec{\ell}^{\,'}\cdot{\vec{s}}^{\,'}+\vec{\ell}\cdot\vec{\ell}^{\,'}}
e^{-{\pi R^{2} \over 
\alpha'}\sum(\ell_{k}\bar{\tau}_{kl}+\ell_{l}^{'})
({\rm Im\,}\tau)^{-1}_{lj}(\ell_{k}\tau_{jk}+\ell_{j}^{'})}}
}
where $\Lambda_{\vec{s}}^{\vec{s}^{\,'}}(\tau,\bar{\tau})$ is the genus-$g$ 
partition 
function for the uncompactified theory. By using Poisson resummation in 
$\ell_{k}$ and summing over all genera it is possible to
prove that \zz\ is invariant under the $\beta$-duality transformation
(see last reference in \refs\rtdual)
\eqn\bet{
R \longrightarrow {\alpha' \over 2R} \hskip 1cm 
\lambda \longrightarrow {\sqrt{\alpha'} \over \sqrt{2}R}\, \lambda  
.}
A remarkable fact about this transformation is that it differs from 
T-duality \tt\ by a numerical factor; in fact \bet\ is equivalent to perform 
a T-duality transformation on the theory with radius $R'=\sqrt{2}R$.
In general $\beta$-duality will be a symmetry of any heterotic string
for which only even spin structures contribute to the genus-$g$
cosmological constant. 

When we are dealing with a type-II theory, in which target space fermions
arise from both the left and right movers, we need to introduce
one phase for each world-sheet chirality. In that case $\beta$-duality 
might not be a 
symmetry of the theory. This is what happens, for example, in the type-II 
superstring \refs\ratickw, which under $\beta$-duality is mapped at one
loop into itself plus some twisted bosonic theory \refs\ros.

Because of the antiperiodic boundary conditions of the
 space-time fermions,  
\zz\ can be interpreted as the genus-$g$ contribution to the canonical 
free energy of a gas of strings at inverse temperature $\beta=2\pi R$. 
Alternatively,  \zz\ can be viewed as the genus-$g$ 
cosmological constant
of an asymmetric orbifold  compactification which breaks space-time
supersymmetry at zero temperature \refs\rramp\rao\rorb. In any case, 
for string theories with
an infinite tower of massive states these compactifications are only
stable in the large radius limit; when the
compactifications radius is of the order of the string scale, tachyons
may appear in the spectrum rendering correlation functions infrared 
divergent. From the thermal viewpoint this is nothing but the old problem
of the Hagedorn temperature caused by the exponential growth of the
number of on-shell states per mass level. These kind of difficulties are
not present when dealing with theories with a finite number
of propagating degrees of freedom, such as the $(2,1)$ string
backgrounds considered in this paper. 

In spite of having no Hagedorn-like interpretation in target space,
 the partition function 
of (2,1) models can be afflicted from similar
 world-sheet instabilites due to 
some states becoming massless at some point of the moduli space of
the toroidal compactification (the off-shell tachyons).
 The only source for such a behavior in 
this case is the coupling of the left-moving NS ground state with the 
winding/momentum states. By going from the winding representation \z\ to 
the mixed
momentum-winding representation it is possible to write the whole partition
function as the sum of contributions coming from the four conjugacy classes of 
$SO(8)$. In the (R,R) sector the phases are equal to one and the
compactification preserves supersymmetry, so we will only have the 
contributions from  
the vector and spinorial conjugacy classes. In the remaining sectors, on 
the contrary, supersymmetry is broken and all the four conjugacy classes
will contribute, in particular the scalar one containing the 
NS tachyonic ground state. Following the analysis of ref. \rorb\ we find 
that, whenever we have antiperiodic boundary conditions in the cycle with
radius $R_{i}$, we will have a potentially divergent term which, for 
example, in the (NS,$*$) sector is of the form
$$\eqalign{
I_{\rm div}&=2\int_{1}^{\infty}{d\tau_{2}\over \tau_{2}} 
\sum_{n_i,\ell_i} \exp{\left\{
-\pi\tau_{2}\left[\left(\ell_1+{1\over 2}\right)^{2}(2R_{1})^{2}+
{\left(n_1+{1\over 2}\right)^{2} 
\over R_{1}^{2}}+\ell_{2}^{2}R_{2}^{2}+{n_{2}^{2}\over 
R_{2}^{2}}-1\right]\right\}} \cr
&\times \int_{-{1\over 2}}^{1\over 2} d\tau_{1}\exp{ 
\left\{ 2\pi i\tau_{1}\left[
2\left(\ell_1+{1\over 2}\right)\left(n_1+{1\over 2}\right)-\ell_{2}n_{2}-
{1\over 2}\right]\right\}.}
}$$
By integrating over $\tau_{1}$ we get a constraint over the integers $n_i$,
$\ell_i$. It is easy to solve this constraint and find  
\eqn\ei{
I_{\rm div}=4\int_{1}^{\infty}{d\tau_{2}\over \tau_{2}}
e^{-{\pi\tau_{2}\over 4}\left(
2R_{1}-{1\over R_{1}}\right)^{2}} 
=4\,E_{1}\left({\pi\over 4}\left|2R_{1}-{1\over R_{1}}\right|^{2}\right)
,}
plus other terms that are regular as functions of $R_{1}$. $E_{1}(z)$ is 
the first exponential integral function whose power expansion around $z=0$ is 
$$
E_{1}(z)=-\gamma_{E}-\log z-\sum_{n=1}^{\infty}{(-1)^{n}z^{n}\over n n!}
,$$
and we get a logarithmic singularity for $I_{\rm div}$ 
in the limit $R_{1}\rightarrow 1/\sqrt{2}$. 
 Moreover, since 
$E_{1}(z)$ is not  singled valued as a function of $z$, 
 we have to introduce 
the absolute value in \ei\ in order to preserve 
the invariance of the original integral under $R_{1}\leftrightarrow 1/2R_{1}$.
In the (NS,NS) contribution we will have a term \ei\ for each radius, so 
divergences will appear when either $R_{1}$ or $R_{2}$ reach the self-dual 
value under $\beta$-duality, $R=1/\sqrt{2}$.

Let us  close this Appendix with some remarks about the equivalence
between the integral over $\CF$ of the partition function and the $\CS$
integral in which we have the 
projection $\Lambda^{s'}_{s}\rightarrow Q^{s'}_{s}$. 
If we were only compactifying one dimension with NS boundary conditions
(i.e. the standard finite temperature situation), 
it would be easy to prove that
both the $\CF$ and the projected $\CS$ representation are equal whenever
the compactification radius satisfy $R>1/\sqrt{2}$. The reason is the 
following; in including the number of propagating fields in a Schwinger
representation of the partition function we can write the number of bosons
and fermions either as a couple of integers $(Q^{0}_{0}+Q^{0}_{1\over 2})
\int_{-1/2}^{1/2}d\tau_{1}$
 and 
$-Q^{1\over 2}_{0}\int_{-1/2}^{1/2}d\tau_{1}$ or in terms of their integral 
representations \qs.
This will provide us with two different integrals over $\CS$, 
$I_{\rm proj}(R)_{\CS}$
and $I(R)_{\CS}$  giving the same function of $R$ if we integrate
first in the real part of $\tau$.  
If we decide to rewrite $I_{\rm proj}(R)_{\CS}$ and $I(R)_{\CS}$ as 
integrals over the fundamental domain $\CF$ we will have to be very careful
with the manipulations involved in the 
coset extension (formula \coset). 
The result will be that both integrals will coincide only when $R>1/
\sqrt{2}$; in particular the extension of $I_{\rm proj}(R)_{\CS}$
will be a regular function for all values of $R$,
 whereas the modular invariant
version of $I(R)_{\CS}$ will have $\beta$-duality and 
 a finite discontinuity in its first derivative 
at the self-dual radius, $R=1/\sqrt{2}$
(cf. \refs\rosva).

This situation changes when we have two compactified dimensions. To clarify
the discussion let us focus in the (NS,R) sector in which we go to the 
$\CS$ representation by disentangling the windings in the cycle with 
NS boundary conditions. Now, because
of the presence of winding modes around the second cycle, 
the projected and unprojected integrals over the strip will not be
equivalent, since
\eqn\ineq{Q^{s'}_{s}\int_{-{1\over 2}}^{1\over 2}d\tau_{1}\,
\sum_{\{\ell\}}e^{-{\pi R_{2}^{2}\over \tau_{2}}
|\ell_{2}\tau+\ell_{2}'|^{2}}
\neq \int_{-{1\over 2}}^{1\over 2} d\tau_{1}\,\Lambda^{s'}_{s}(\tau)
\sum_{\{\ell\}}e^{-{\pi R_{2}^{2}\over \tau_{2}}|\ell_{2}\tau+\ell_{2}'|^{2}}.
}
Therefore, the corresponding modular invariant extensions of both
results will only be equivalent in the limit in which we decompactify 
the second cycle $(R_{2}\rightarrow \infty)$, and keep the first radius
above $1/\sqrt{2}$. It is however important to
notice that the left-hand
side of \ineq\ is recovered inside the right-hand side as the zero mode
part in the expansion of $\Lambda^{s'}_{s}$ in powers of $q$ and $\bar{q}$.
This suggests that the {\it full} integral over the fundamental domain
$\CF$ contains the projected integrals \res, plus other terms that are singular
at some codimension one regions of the moduli space of $T$ and $U$. The nature
of these logarithmic singularities, analyzed above, supports the view that 
it is only the projected piece of the integral that factorizes into a $T$ and 
a $U$-dependent part, whereas this would not be the case for the terms that 
we are dropping in the projection $\Lambda^{s'}_{s}\rightarrow Q^{s'}_{s}$.

\listrefs
\bye